\documentclass{pasj00}
\draft

\begin{document}
\SetRunningHead{Author(s) in page-head}{Running Head}
\Received{2004/11/11}
\Accepted{2004/12/27}

\title{Black Hole Shadows of Charged Spinning Black Holes}

\author{Rohta \textsc{Takahashi} %
}
\affil{Yukawa Institute for Theoretical Physics,
Kyoto University, Sakyo-ku,  Kyoto 606-8502}
\email{rohta@yukawa.kyoto-u.ac.jp}


%


\maketitle

\begin{abstract}
We propose a method for measuring the black hole charge by imaging 
a black hole shadow in a galactic center by future interferometers. 
Even when the black hole is uncharged, 
it is possible to confirm the charge neutrality by this method. 
We first derive the analytic formulae of the black hole shadow in an optically 
thin medium around a charged spinning black hole, and then investigate how contours of 
the black hole shadow depend on the spin and the charge of the black hole 
for several inclination angles between the rotation axis of the black hole and the observer.  
This method only assumes stationary black hole and general relativity.  
By fitting the formula of the contours of 
the shadow to the observed image of the shadow,   
in addition to the black hole charge, one can also determine the black hole 
spin and the inclination angle 
without any degeneracy among the charge, the spin, and the inclination angle 
unless the inclination angle is null. 
If the maximum width of the shadow smaller than $4(1+2^{1/2})M$ 
or the minimum width of the shadow smaller than $9M$ are measured observationally, 
we can safely conclude that the black hole is charged. 
Here, $M$ is the gravitational radius, i.e. the half of the Schwarzschild radius. 
\end{abstract}

\section{Introduction}
According to the no--hair theorem of a black hole 
(e.g. Ruffini, Wheeler 1971; Frolov, Novikov 1998), 
or equivalently the uniqueness theorem of a black hole (e.g. Hawking, Ellis 1973; Wald 1984), 
the parameters of a stationary black hole are the mass, $M$, the angular momentum, 
$S$, and the charge, $Q$. 
Observational determinations of all the parameters of a black hole is one of the 
greatest challenges in astrophysics. 
The masses of black holes were actually measured observationally during the last 
century (e.g. Miyoshi et al. 1995). 
Also, black hole spins are proposed for several massive or stellar-mass black hole 
candidates 
(e.g. Bromley et al. 1998; Melia et al. 2001; Miller et al. 2002; Genzel et al. 2003a). 
On the other hand, compared with the mass and the spin, 
the charge has not been observationally measured so far. 
This is partly because there is no established method to measure the black hole charge. 
Theoretically, some authors have considered that matter around a black hole is 
neutral in charge, and that a black hole in the real universe has no charge. 
It seems to be probable that even if a black hole has some charge, 
it will positively swallow 
any inversely charged particles and then move into the neutral state. 
Moreover, for a black hole with a mass smaller than $\sim10^8 M_\odot$, 
it is well-known that 
the maximum charge of the black hole is prohibited by the spontaneous loss of charge 
by black holes (Gibbons 1975). 
On the other hand, for example, 
if a black hole is surrounded by a plasma consisting of electrons and 
protons in magnetic fields, some electrons and protons may move differently,  
escaping from the electron--proton coupling 
because of a different inertial force. 
Some authors have actually 
proposed several mechanisms to oscillate between positive and negative 
charge or to induce an electric charge 
into a black hole (Wald 1974, 1984; Levich et al. 1972; de Diego et al. 2004). 
At least, observationally, 
the neutrality of black holes in the real universe has not been confirmed so far. 
In the present paper we propose a simple method to determine the black hole charge 
by observing the black hole shadow. 
Imaging the black hole shadow is one of the observational targets of future interferometers, 
such as the VSOP-2 mission (Hirabayashi et al. 2001) or the MAXIM projects (Cash et al. 2000). 
In the present work, we first derive an analytic formula of the contour of a black hole shadow 
in an optically thin medium in the Kerr--Newman metric;   
based on this formula we then investigate the parameter dependence of the black hole 
shadow, which depends on the mass, $M$, the angular momentum, $S$, and the charge, $Q$, 
of the black hole, and the inclination angle, $i$, between us and the rotation axis of 
the black hole. 

\section{Analytical Formula of the Shadow of Charged-Spinning Black Hole}

We assume the background metric of the Kerr--Newman geometry written by the 
Boyer--Lindquist coordinates with $c=G=1$, 
\begin{eqnarray}
ds^2=-e^{2\nu}dt^2+e^{2\varphi}(d\phi-\omega dt)^2+e^{2\mu_1}dr^2+e^{2\mu_2}d\theta^2.  
\end{eqnarray}
Here, 
$e^{2\nu}\equiv\Sigma\Delta/A$,  
$e^{2\varphi}\equiv\sin^2\theta A/\Sigma$,  
$e^{2\mu_1}\equiv\Sigma/\Delta$,  
$e^{2\mu_2}\equiv\Sigma^2$, 
and $\omega=(2Mr-Q^2)a/A$,  
where $\Delta\equiv r^2-2Mr+a^2+Q^2$, $\Sigma\equiv r^2+a^2\cos^2\theta$, 
$A\equiv (r^2+a^2)^2-a^2\Delta\sin^2\theta$, 
and $a\equiv S/M=$angular momentum per unit mass, i.e. black hole spin. 
The Kerr--Newman geometry has a horizon, and describes a black hole,  
if and only if $M^2\ge Q^2+a^2$.  
The event horizon, $r_{{\rm h}+}$, is located at 
$r_{{\rm h}+}\equiv M+(M^2-Q^2-a^2)^{1/2}$. 
The $r$ and $\theta$ components of null geodesics are 
described as (Carter 1968; Misner et al. 1973) 
$\Sigma dr/d\lambda=R^{1/2}$ and $\Sigma d\theta/d\lambda=\Theta^{1/2}$. 
Here, $\lambda$ is an affine parameter, 
\begin{equation}
R=[E(r^2+a^2)-L_za]^2-\Delta[(L_z-aE)^2+\mathcal{Q}], 
\end{equation}
and 
\begin{equation}
\Theta=\mathcal{Q}-\cos^2\theta(-a^2E^2-L_z^2/\sin^2\theta), 
\end{equation}
where $E$, $L_z$, and $\mathcal{Q}$ are the energy at infinity, 
the axial component of the angular momentum, and Carter's constant, respectively. 
As in Chandrasekhar (1983), it is convenient to minimize the number of parameters 
by letting $\xi=L_z/E$ and $\eta=\mathcal{Q}/E^2$. 
The equations determining the unstable orbits of 
constant radius are $R=0$ and $\partial R/\partial r=0$ 
(e.g. Chandrasekhar 1983; Bardeen 1973; Young 1976).  
In the case of a Karr--Newman black hole, these equations are reduced to  
\begin{eqnarray}
r^4+(a^2-\xi^2-\eta)r^2+2M[\eta+(\xi-a)^2]r
-a^2\eta-Q^2[\eta+(\xi-a)^2]=0 
\label{R1}
\end{eqnarray}
and
\begin{eqnarray}
2r^3+(a^2-\xi^2-\eta)r+M[\eta+(\xi-a)^2]=0.  
\label{R2}
\end{eqnarray}
From equations (\ref{R1}) and (\ref{R2}), we obtain 
\begin{equation}
\xi=\frac{-r^3+3Mr^2-a^2r-Ma^2-2Q^2r}{a(r-M)}
\label{xi} 
\end{equation}
and
\begin{equation}
\eta=\frac{r^3(-r^3+6Mr^2-9M^2r+4Ma^2)-4r^2Q^2(r^2+a^2+Q^2-3Mr)}{a^2(r-M)^2}. 
\label{eta}
\end{equation}
When $Q=0$ and $a=M$, equations (\ref{xi}) and (\ref{eta}) result in 
equation (238) in section 63 of Chandrasekhar (1983). 
Next, we calculate the relation between two constants, $\xi$ and $\eta$, 
and the celestial coordinates, $x$ and $y$, of the image as seen by an observer 
at infinity, who receives light ray propagating in the Kerr--Newman geometry. 
The tetrad components of the four momentum $[p^{(t)},~p^{(r)},~p^{(\theta)},~p^{(\phi)}]$ 
with respect to locally nonrotating reference frames (LNRFs) can be written as:  
\begin{eqnarray} 
p^{(t)}&=&e^{-\nu}(E-\omega L_z),\\
p^{(r)}&=&-e^{-\mu_1}p_r, \\
p^{(\theta)}&=&-e^{-\mu_2}{p_\theta},\\
p^{(\phi)}&=&-e^{-\varphi}p_\phi, 
\end{eqnarray} 
where $p_r=-R^{1/2}/\Delta$, $p_\theta=-\Theta^{1/2}$, and $p_\phi=-L_z$. 
Here, $R$ depends on the black hole charge, $Q$. 
By using these equations, we can calculate $x$ and $y$ as follows: 
\begin{eqnarray}
x&=&\lim_{r\to \infty}\left[\frac{-rp^{(\phi)}}{p^{(t)}}\right]
=-\frac{\xi}{\sin{i}},\label{x}\\
y^2&=&\lim_{r\to \infty}\left[\frac{rp^{(\theta)}}{p^{(t)}}\right]^2
=\eta+a^2\cos^2{i}-\frac{\xi^2}{\tan^2{i}},\label{y} 
\end{eqnarray}
where $i$ is the inclination angle between the observer and the direction of 
the rotation axis of the black hole. 
Since we take the limit $r\to\infty$, the relations between $\eta$, $\xi$, 
and $(x,~y)$ become the same as those in the Kerr metric (Chandrasekhar 1983). 
By using equations (\ref{xi}), (\ref{eta}), (\ref{x}), and (\ref{y}), 
we can draw the contour of the black hole shadow in the $(x,~y)$ plane. 
The condition $y^2\ge 0$ along with equation (\ref{y}) determines 
the range of $r$ of equations (\ref{xi}) and (\ref{eta}). 
It is noted that the contour of the black hole shadow is symmetric with respect to 
the $x$-axis. 
It is clear that there is no degeneracy between the charge, $Q$, 
the spin, $a$, and the inclination angle, $i$, in this formula of a black hole 
shadow.  
Since black hole shadows depend on the black hole charge, $Q$, in forms of $Q^2$, 
these 
black hole shadows are symmetric with respect to a change of the sign of the black hole 
charge. 

\section{Qualitative and Quantitative Analyses of the Shadow}

In figure 1, we show examples of these black hole shadows for 
$Q/M=0$ (thin solid lines), $|Q|/M=0.7$ (dashed lines) and the 
maximum Kerr--Newman black hole, i.e. $a^2+Q^2=M^2$ (thick solid lines). 
Although some of the similar contours were calculated by Young (1976) in a study of 
the cross section of gravitational capture by black holes, 
he did not present an analytic formula, such as equations  
(\ref{xi}) and (\ref{eta}). 
For a fixed non-zero spin, $a\neq 0$, the shapes of the black hole 
shadow are more elongated for larger inclination angles 
due to the black hole rotation. 
This feature can be clearly seen in the panels for $a/M=0.999$ in figure 1. 
Also, for a fixed non-zero inclination angle, 
$i\neq 0^\circ$, the shapes of the black hole 
shadow have more elongated shapes for larger black hole spins. 
In addition to the elongation features, for non-zero inclination angles, 
$i\neq 0^\circ$, the position of the black hole shadow is shifted from the rotation axis 
of the black hole. 
This feature due to the black hole rotation 
has already been pointed out by Takahashi (2004) for the black hole shadows in accretion disks. 
As can clearly be seen in figure 1, 
in the case of $|Q|/M\neq 0$, for a fixed black hole spin, 
the size of the black hole shadow become smaller than the black hole shadow with $Q/M=0$. 
This is because the event horizon, $r_{{\rm h}+}=M+(M^2-a^2-Q^2)^{1/2}$, is 
smaller for a larger $|Q|$. 
In addition to the smallness feature, for a fixed non-zero black hole spin and 
a non-zero inclination angle, 
the black hole shadow has more elongated shapes for a larger $|Q|$. 
This is due to the black hole rotation, as mentioned above. 
The angular momentum of the frame dragging, $\omega$, in the Kerr--Newman black hole 
is expanded in the equatorial plane as 
\begin{equation}
\omega=\frac{2Ma}{r^3}-\frac{2Ma^3-Q^2a}{r^4}+O\left(\frac{1}{r^5}\right). 
\end{equation}
Because the frame--dragging effect decreases in proportion to $a/r^3$, as shown here, 
the frame--dragging effects on the propagation of photons become more significant 
for more inner regions, i.e. a smaller event horizon.  
Therefore, the black hole shadow has more elongated shapes for a larger $|Q|$, i.e. 
a smaller event horizon, $r_{{\rm h}+}$. 

Since the black hole shadow is symmetric with respect to the $x$-axis, 
the maximum size of black hole shadows, $\Delta_{\rm max}$, is 
$\Delta_{\rm max}=2y_{\rm max}$. 
The minimum size, $\Delta_{\rm min}$, 
of the black hole shadow is $\Delta_{\rm min}=x_{\rm max}-x_{\rm min}$.  
Here, $y_{\rm max}$, $x_{\rm max}$, and $x_{\rm min}$ are the maximum 
$y$-coordinate, 
the maximum $x$-coordinate, and the minimum $x$-coordinate of the contour of 
the black hole shadow, respectively. 
After some calculations, 
the ranges of $\Delta_{\rm max}$ and $\Delta_{\rm min}$ of the black hole shadows 
including the charged spinning black hole are calculated as follows: 
\begin{eqnarray}
8M&\le&\Delta_{\rm max}\le6\times3^{1/2}M, \\
8M&\le&\Delta_{\rm min}\le6\times3^{1/2}M. 
\end{eqnarray}
Here, $6\times3^{1/2}M\sim10.3923M$. 
The maximum $\Delta_{\rm max}$ and the maximum $\Delta_{\rm min}$ 
are realized for $Q/M=0$ and $a/M=0$. 
The minimum $\Delta_{\rm max}$ and the minimum $\Delta_{\rm min}$ 
are realized for $|Q|/M=1$. 
In the case of an uncharged black hole, i.e. $Q=0$, the ranges of 
$\Delta_{\rm max}$ and $\Delta_{\rm min}$ are as follows: 
\begin{eqnarray}
4(1+2^{1/2})M&\le&\Delta_{\rm max}\le6\times3^{1/2}M, \\
9M&\le&\Delta_{\rm min}\le6\times3^{1/2}M. 
\end{eqnarray}
Here, $4(1+2^{1/2})M\sim9.6568M$. 
The minimum $\Delta_{\rm max}$ is realized when $a/M=1$ and $i=0^\circ$. 
The minimum $\Delta_{\rm min}$ is realized when $a/M=1$ and $i=90^\circ$. 
Note that the minimum $\Delta_{\rm max}$ and the mimimun $\Delta_{\rm min}$ 
of a maximally charged black hole, i.e. $|Q|/M=1$, 
are different from those for uncharged black holes. 
Thus, even when the subtle shape of the black hole shadow can not be observed because of 
an insufficient spatial resolution of a telescope, 
if a $\Delta_{\rm max}$ smaller than $4(1+2^{1/2})M$ or 
a $\Delta_{\rm min}$ smaller than $9M$ is measured observationally, 
we can safely conclude that the black hole is charged.  

\begin{figure*}
\includegraphics[scale=1.4]{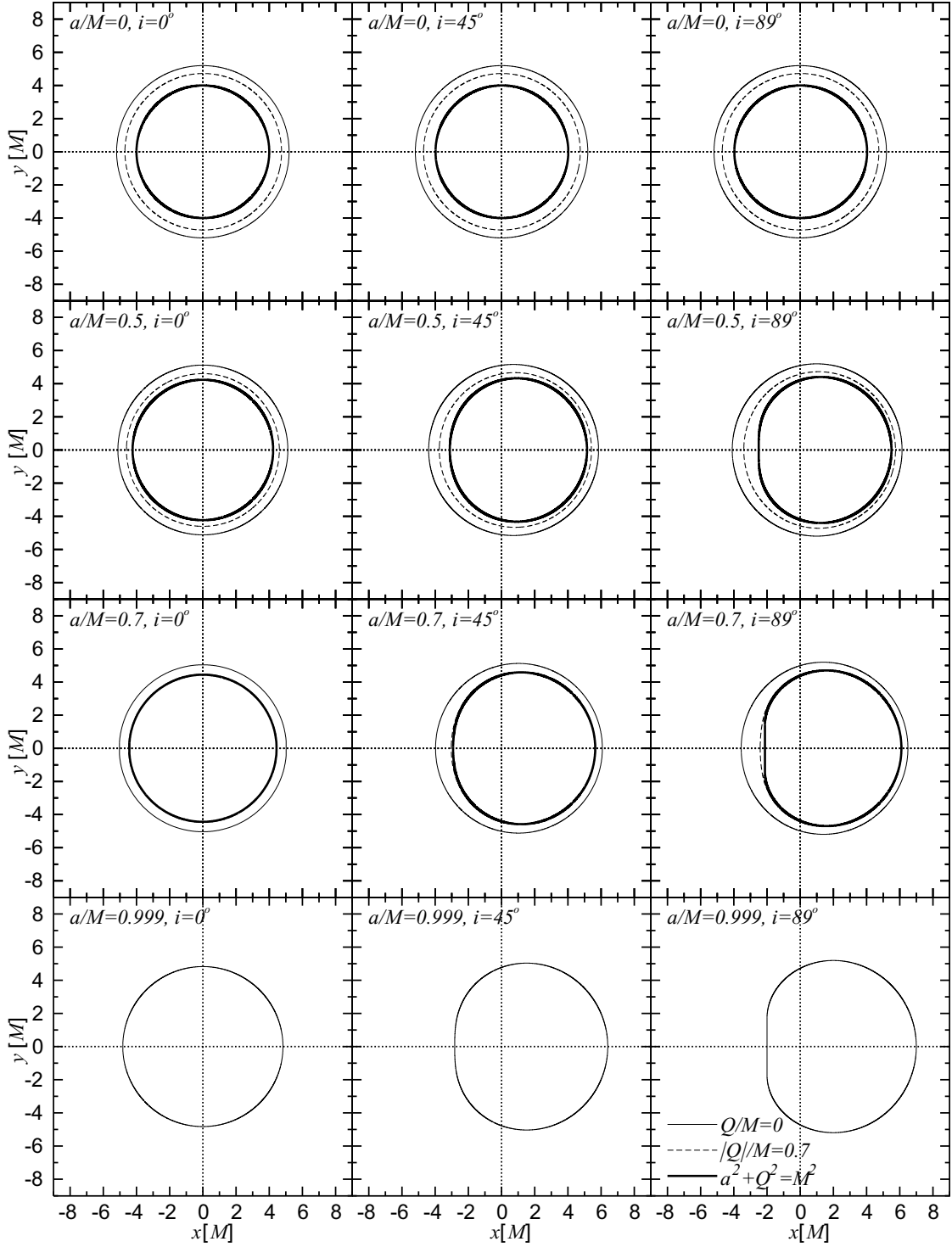}
\caption{
Examples of black hole shadows in an optically thin emitting medium for several 
combinations of spin, $a$, and charge, $Q$, with $a^2+Q^2\le M^2$. 
See the text for an explanation. 
\label{BHS}
}
\end{figure*}

\section{Discussion}
When we can have observational facilities with infinite spatial resolution, and 
the subtle shapes of black hole shadows can be clearly measured observationally, 
we can clearly observe the black hole shadow; the spin $a/M$, the charge $Q$, and the inclination 
angle $i$ can be determined  
by fitting the contours of the black hole shadows calculated by equations  
(\ref{xi}), (\ref{eta}), (\ref{x}), and (\ref{y}) to the observed image of 
the black hole shadow if the black hole mass is determined by other methods. 
If such a subtle observation can not be made, roughly speaking,  
the three parameters of the spin, $a/M$, the charge, $Q$, and 
the inclination angle, $i$, can be measured from the three quantities of 
the maximum width and the minimum width of the black hole, and 
the shift, $\delta$, of the black hole shadow from the mass center of the black hole. 
Here, $\delta$ is defined as $\delta=(x_{\rm max}+x_{\rm min})/2$. 
When the spatial resolutions of the observational facilities are not sufficient to 
image the subtle shapes of the black hole shadows, 
in order to determine $a/M$, $Q$, and $i$ by observations of black hole shadows 
without any degeneracy among these parameters 
it is crucially important to determine  
the positions of the mass center of the black hole in the black hole shadow 
precisely in order to measure $\delta$. 
In the case of an uncharged spinning black hole, the range of 
$\delta$ is $0\le\delta\le 2.5M$. 
The maximum shift, $\delta=2.5M$, is realized when $a/M=1$ and $i=90^\circ$. 
In the case of Sgr A*, 
the mass of the central black hole can be measured by observations 
of the stellar motion around the black hole (Genzel et al. 2003b).  
From these stellar motions, we can determine the position of 
the mass center of the black hole and the black hole spin (Cunningham, Bardeen 1973) 
in principle. 
Also, several astrometry satellite missions with a positional accuracy of 10 $\mu$as  
which will provide plenty of data about the stellar motions around Sgr A*, are planned 
(Belokurov, Evans 2002; Gouda et al. 2002).  
There is also a simple additional method using the brightness distribution of the accretion 
matter to determine the black hole mass center (Takahashi 2004). 
\ \newline

The author is grateful to Professor Shin Mineshige for continuous encouragements. 
This work is supported by the Grant-in-Aid  for the 21st Century COE 'Center for Diversity 
and Universality in Physics'. 
The author is supported by a Research Fellowship of the Japan Society for the Promotion 
of Science for Young Scientists. 


\end{document}